\documentclass[12pt]{article}
\usepackage{amsmath}
\usepackage{graphicx}
\usepackage{enumerate}
\usepackage{natbib}
\usepackage{url} % not crucial - just used below for the URL 
\usepackage{ulem}
\usepackage{amsfonts}
\usepackage{amssymb}
\usepackage{bm}
\usepackage{color}
\usepackage{hyperref}
\usepackage{keyval}
\usepackage{multirow}

\newcommand{\blind}{0}

\newcommand{\real}{{\mathbb R}}
\newcommand{\reald}{\real^{\rm{d}}}
\newcommand{\realp}{\real^{\rm{p}}}

\newcommand{\pbty}{\mathbb P}

\begin{document}

\def\spacingset#1{\renewcommand{\baselinestretch}%
{#1}\small\normalsize} \spacingset{1}

%%%%%%%%%%%%%%%%%%%%%%%%%%%%%%%%%%%%%%%%%%%%%%%%%%%%%%%%%%%%%%%%%%%%%%%%%%%%%%
\if0\blind
{
  \title{\bf  Estimation for  bivariate  quantile varying coefficient model}
  \author{Linglong Kong$^*$, Haoxu Shu, Giseon Heo\thanks{
    The authors gratefully acknowledge Natural Sciences and Engineering Research Council of Canada (\textit{NSERC}).  Heo's parimary affiliation  is in School of Dentistry, University of Alberta and is cross appointed in the   Department of Mathematical and Statistical Sciences.
Heo also  acknowledges \textit{McIntyre Memorial Funding}  in  Orhodontics division, School of Dentistry. } \\
   Department of Mathematical and Statistical Sciences,  \\ University of Alberta\\
   Edmonton, AB T6G 2G1 Canada\\
    and \\
    Qianchuan He  \\
    Public Health Sciences Division, \\Fred Hutchinson Cancer Research Center, \\Seattle, WA 98109 USA
}
  \maketitle
} \fi

\if1\blind
{
  \bigskip
  \bigskip
  \bigskip
  \begin{center}
    {\LARGE\bf Title}
\end{center}
  \medskip
} \fi

\bigskip
\begin{abstract}
We propose a bivariate  quantile regression method for the bivariate varying coefficient model through a directional approach.
The varying coefficients are approximated by the B-spline basis and an $L_{2}$-type penalty is imposed to achieve desired smoothness. We develop a multistage estimation procedure based the Propagation-Separation~(PS) approach to borrow information from nearby directions. The PS method is capable of handling the computational complexity raised by simultaneously considering multiple directions to efficiently estimate varying coefficients while guaranteeing certain smoothness along directions. We reformulate the optimization problem and solve it by the Alternating Direction Method of Multipliers~(ADMM), which is implemented using R while the core is written in C to speed it up. Simulation studies are conducted to confirm the finite sample performance of our proposed method. A real data on Diffusion Tensor Imaging~(DTI) properties from a clinical study on neurodevelopment  is analyzed.
\end{abstract}

\noindent%
\vfill

\newpage
\spacingset{1.45} % DON'T change the spacing!

\section{Introduction}\label{intro}
Consider a regression model 
$Y_i=f(X_i, \beta)+ \epsilon_i,$ for $i=1, \ldots, n,$ 
where $Y_i$ represents the response, $X_i$ is a vector of covariates,  $\beta$ is the regression coefficient parameter, $f$ is a known or unknown function of the covariates $X_i$ and $\beta$, and $\epsilon_i$ is the error term. Assuming a known function $f$, in the ordinary least square (OLS) regression, the parameter  $\beta$ is estimated by minimizing the sum of squared residuals $\sum_{i=1}^{n}(y_{i}-f(\bm{x}_{i},\beta))^2.$ While in quantile regression (QR),  the QR effect $\beta_{\tau}$ for $\tau \in (0,1)$  is obtained by minimizing
\begin{equation} \sum_{i=1}^{n}\rho_{\tau}(y_{i}-f(\bm{x}_{i},\beta)),
\label{qreg}
\end{equation}
where $\rho_{\tau}(u)=u(\tau-I(u<0))$ is the quantile loss function \citep{koenker1978regression}. When  $\tau=.5$, the quantile regression becomes the least absolute deviation (LAD) regression, an alternative to the OLS,  which estimates the conditional median instead the conditional mean of the response. 
By assuming that the error follows the asymmetric Laplace distribution, the maximum likelihood approach or Bayesian method can be used to estimate $\beta$ \citep{yu2001bayesian,yuan2010bayesian,yang2015posterior}. 

There are at least three advantages to consider conditional quantiles instead of the conditional mean in a regression setting. 
First, quantile regression, in particular median regression, provides an alternative and complement 
to mean regression while being resistant to outliers in responses; 
in addition, quantile regression is more efficient than mean regression when the error follows a distribution with heavy tails. 
Second, quantile regression is capable of dealing with heteroscedasticity, the situation in which variances depend on certain covariates.
 More importantly, quantile regression can give a more complete picture on how the responses are affected by covariates, 
particularly the tail behavior of the response conditional on covariates, for example in economic and actuaries. For more background on quantile regression, 
see the monograph by \cite{koenker2005quantile}.

Inspired by the success of the univariate quantile (for a single response), researchers began to extend univariate quantiles to multivariate quantiles (i.e., for multiple responses). For example, the definition of multivariate quantile proposed by  \cite{chaudhuri1996geometric} is a generalization of what was proposed by \cite{koenker1978regression} in univariate cases. In general, there is an associated multivariate median for any concept of multivariate quantiles. In other words, we first solve for the multivariate median and then extend it to multivariate quantiles. There are many methods to define a multivariate median, for more details see \cite{small1990survey}. Some methods for multivariate medians  can be extended to multivariate quantiles, and different authors gave distinct multivariate quantile extensions, such as the spatial $L_1$ median  \cite{koltchinskii1996m} and the Oja median computed by \cite{ronkainen2003computation}. However, not every multivariate extension of quantiles can be obtained from a multivariate median. Alternatively, multivariate quantiles can be generalized directly from univariate quantiles (via approaches based on norm minimization), such as the one derived by \cite{abdous1992note}. 

The methods for  univariate quantile regression  are, in general, not easily applicable to multivariate quantile regression because of  many reasons, one of which is the difficulty of interpretation.
Nevertheless, multivariate quantiles via a directional approach is one of the successful extensions \citep{breckling1988m,wei2008approach,hallin2010multivariate,kong2012quantile}. This approach was proposed and applied to bivariate growth charts by \cite{kong2012quantile}. The authors showed that the analysis in terms of directional quantiles and their envelopes offers a straightforward probabilistic interpretation and thus conveys a concrete quantitative meaning. They  also demonstrated  that the directional quantile regression can facilitate the construction of bivariate growth charts and provide richer information than univariate growth charts. 

Borrowing the ideas and notations from \cite{kong2012quantile}, we explain directional  quantile regression as follows.
Let $\bm{S}^{\rm{d-1}}=\{\bm s\in \reald:\|\bm s\|=1\}$ and $X$ be a random vector in $\reald$ with distribution $\pbty$. Given $\bm s\in\bm{S}^{\rm{d-1}}$ and $0<\tau<1$, the $\tau$-th directional quantile
in the direction $\bm s$ is defined as the $\tau$-th quantile of the corresponding projection of the distribution of $X$, that is,

\begin{equation}\label{q_df}
	Q(\tau,\bm s)=Q(\tau,\bm s^{T}X)=\inf\{u:\pbty(\bm s^{T}X \leq u)\geqslant \tau\}.
\end{equation}
For fixed $\tau\in(0,1/2]$, the $\tau$-th directional quantile envelope generated by $Q(\tau,\bm s)$ is defined as the intersection of halfspaces,
\begin{equation}\label{qe_dfn}
	D(\tau)=\bigcap_{\bm s\in\bm{S}^{\rm{d-1}}}H(\bm s,Q(\tau,\bm s)),
\end{equation}
where $H(\bm s,q)=\{x:\bm s^{T}x\geqslant q\}$ is the supporting halfspace determined by $\bm s\in\bm{S}^{\rm{d-1}}$ and $q\in\real.$
This multivariate directional quantile concept can be easily extended to multivariate directional quantile regression \citep{kong2009multivariate,kong2012quantile}. In general, given a direction $\bm s$ and a multivariate regression model  
	$\bm Y\sim \bm f(\bm X,\bm{\beta}),$	
where $\bm Y=(Y_{1}, \ldots, Y_{d})^T$, we project $\bm Y$ on the direction $s$  and denote it as $\bm Y^{*}=\bm s^{T}\bm Y.$
Then using data $\{(\bm y_{i}^{*},\bm x_{i})\}, i=1,...,n$, for $0<\tau<1$, we can obtain the $\tau$-th quantile by minimizing $\sum\rho_{\tau}\left(\bm y_{i}^{*}-\bm f(\bm x_{i},\bm{\beta}_{\tau})\right)$, where $(\bm y_i, \bm x_i)$ are observations from $(\bm Y,\bm X)$. Similarly, we can have directional quantiles at any other selected directions. Using these quantiles, we can generate the $\tau$-th conditional directional quantile envelope of $\bm Y$ for any given $\bm X$.  The directional quantile envelopes are essentially Tukey's depth contours \citep{tukey1975mathematics,kong2010smooth} and the directional quantile regression envelopes are herein the conditional Tukey's depth contours. The directional quantile regression inherits good properties from Tukey's depth and also provides straightforward probabilistic interpretation; for more details, see \citep{kong2009multivariate,kong2012quantile}.  

Since the varying coefficient model was systematically introduced by \cite{hastie1993varying}, it rapidly becomes a powerful statistical tool for time series and longitudinal data \citep{wu1998asymptotic, huang2002varying,fan2003adaptive}. Recently, it has been developed for functional data analysis; see \citep{silverman2005functional,zhang2007statistical,zhu2012multivariate}.  In quantile regression literatures, there are also many new developments, for example, \cite{honda2004quantile, kim2007quantile, cai2008nonparametric, wang2009quantile,tang2013variable}, and \citet{zhao2013variable}, just to name a few. However, in the functional data anaysis framework, there are only limited methodologies available, for example, \cite{Zhou2015qr} developed a novel method for quantile regression with varying coefficients for univariate functional responses based on the local polynomial kernel smoothing technique. In real world, however, multiple measurements may be taken along a series of spatial or temporal points. For example, in diffusion tensor imaging (DTI) studies multiple fiber measurements like fractional
anisotropy (FA) and mean diffusivity (MD) are measured along major fiber tracts \citep{zhu2010multivariate,zhu2011fadtts,zhu2012multivariate}. To jointly model the multiple functional responses with the spatial positions will enhance the strength shared among the responses and positions, and thus will improve the efficiency of the quantile estimates and provide more information to reveal some underlying truth that can not be obtained by individual modeling. Unfortunately, there is no existing method in the literature that can handle such a task in the quantile regression with varying coefficients for functional responses.

In this article, we propose a novel estimation procedure in bivariate quantile varying coefficient model for functional responses 
to investigate the association between the responses and the covariates of interests, such as gender and gestational age \citep{zhu2011fadtts}. 
Our estimation method is based on the directional quantile concept and has the following innovative features. First, by jointly modeling the bivariate functional responses our method provides a conditional bivariate quantile envelope tube along the spatial or temporal positions which is capable of uncovering the underlying information that can not be obtained by univariate quintile regression. Second, to achieve the desired smoothness along the spatial or temporal positions for the quantile envelope tubes, an $L_2$-type roughness penalty is imposed to estimate the varying coefficients approximated by B-splines \citep{koenker2005quantile}. Third, to improve the efficiency of the quantile estimates, we develop 
a multistage estimation procedure based on the propagation-separation~(PS) approach \citep{polzehl2000adaptive, polzehl2006propagation} to gradually borrow information from nearby directions with increasing number of directions. 
 The PS method is capable of handling the computational complexity raised by simultaneously considering multiple directions to efficiently estimate varying coefficients while guaranteeing certain smoothness along directions. To the best of our knowledges, this is the first method to construct the directional quantile regression envelopes 
by simultaneously considering multiple directions. Forth, we reformulate the optimization problem and solve it by the Alternating Direction Method of Multipliers~(ADMM) repopularized by \cite{boyd2011distributed}, which is implemented using R while the core is written in C to speed it up. ADMM is efficient to tackle our optimization problem with a nonsmooth quantile loss function plus a $L_2$ type penalty.

We organize our article as follows.  
We introduce the  bivariate quantile varying coefficient model and define the objective function with penalties by approximating the varying coefficients using B-splines in Section \ref{method}.
We adapt the PS approach to our estimation and describe the multistage  estimation procedures for varying coefficients in Section \ref{estimate}.
In Section \ref{admm}, we reformulate the optimization problem and solve it by ADMM. 
Our proposed estimation procedure and algorithm are examined in the simulation studies in Section \ref{sim}.
As a demonstration, fractional
anisotropy (FA) and mean diffusivity (MD) along the genu fiber bundle of
 the corpus callosum (GCC) of  the diffusion tensor imaging (DTI)  from a clinical study on neurodevelopment are analyzed in Section \ref{real}.
We summarize our results and discuss future work in Section \ref{conclusion}. 

\section{Methodology}

\subsection{Bivariate quantile varying coefficient model}\label{method}
Motivated by the generalized regression quantiles of \cite{guo2015functional} and the multivariate varying coefficient model in \cite{zhu2011fadtts}, we propose our bivariate quantile varying coefficient model (BQVCM) through a directional quantile approach \citep{kong2009multivariate,kong2012quantile}. 
 Let $\bm{Y}(t)=(Y_{1}(t),Y_{2}(t))^T\in\real^2$ be a bivariate functional response at time $t$, where $t\in [0,1]$, and  $\bm{X}\in \realp$ be the covariates of interest. 
  For fixed $0<\tau<1 $, given a direction $\bm s \in \bm {S}^1$, the bivariate quantile varying coefficient model defines the $\tau$-th directional quantile of $\bm{Y}(t)$ at $t$ given $\bm{X}$, denoted by $Q_{\bm Y(t)|\bm X}(\tau,\bm s)$, 
  \begin{equation}\label{bqvcm}
  Q_{\bm Y(t)|\bm X}(\tau,\bm s) = \bm f(\bm X,\bm \beta_\tau(\bm s, t)), 
  \end{equation}
 where $\bm f$ characterizes the dependency of the quantile of $\bm Y(t)$ on $\bm X$ and the varying coefficients $\bm \beta_\tau(\bm s, t)$ are the $\tau$-th quantile parameters to be estimated. 
  To be simple, in this article we assume a linear dependency quantile structure. That is 
    \begin{equation}\label{bqvcm1}
Q_{\bm Y(t)|\bm X}(\tau,\bm s) = \bm X^T\bm \beta_\tau(\bm s, t), 
  \end{equation}
  where $\bm \beta_\tau(\bm s, t) = (\beta_{\tau1}(\bm s, t),\cdots,\beta_{\tau p}(\bm s, t))^T\in \realp$ varies along $t$ and also depends on the direction $\bm s$.  In this article, we always fix a $\tau\in (0,1)$ to estimate $\bm \beta_\tau(\bm s, t)$. Therefore, to simplify the notations we hereafter will drop the subscript $\tau$ if there is no confusion. The $\tau$-th directional quantile envelope generated from the BQVCM \eqref{bqvcm1} can be defined as in \eqref{qe_dfn},
  \begin{equation}\label{qe_dfn1}
	D(\tau,t)=\bigcap_{\bm s\in\bm{S}^{1}}H(\bm s,Q_{\bm Y(t)|\bm X}(\tau,\bm s)),
\end{equation} 
where $D(\tau,t)$ depends on $t$ and essentially constructs a bivariate quantile envelope tube along $t$ for each fixed $\tau$.  Note that the quantile envelope tubes 
$D(\tau,t)$ are nested in terms of $\tau$, as Tukey's depth contours \citep{tukey1975mathematics}. 
  
For given observations $\{\bm y_i(t_j),\bm x_i\}$, where $i=1,\cdots,n$ and $j=1,\cdots,J$.  For $0<\tau<1$,  to estimate $\bm \beta(\bm s, t)$ in \eqref{bqvcm1}, 
we minimize 
\begin{equation}\label{bqsvcm}
\bm L\left(\bm y,\bm x,\bm s;\bm \beta(\bm s)\right)=\sum_{i=1}^{n}\sum_{j=1}^{J}\rho_{\tau}\left(\bm{s}^{T}\bm{y}_{i}(t_{j})-\bm{x}_{i}^T\bm{\beta}(\bm{s},t_{j})\right),
\end{equation}
where $\rho_{\tau}$ is the quantile loss function defined in \eqref{qreg}.  Under certain conditions, the varying coefficients $\bm{\beta}(\bm{s},t_{j})$ can be well approximated by B-splines \citep{huang2002varying,huang2004polynomial}.  Let $\bm H(t) = (h_1(t),\cdots,h_M(t))^T$ be the selected B-spline basis, where $M$ is the total number of basis functions. The coefficients $\bm{\beta}(\bm{s},t)$  are approximated by $\bm C(\bm s)\bm H(t)$ with $\bm C(\bm s)$ being a $p\times M$ coefficient matrix.  Letting $\bm B(\bm s)$ be the vectorization of $\bm C(\bm s)$, i.e., a $pM$ dimensional vector, then $\bm C(\bm s)\bm H(t)=\bm I_p \otimes \bm H^T(t) \bm B(\bm s)$. Then 
the loss function \eqref{bqsvcm} is rewritten as 
\begin{equation}\label{bqsvcm0}
\bm L_b(\bm y,\bm x,\bm s;\bm B (\bm s))=\sum_{i=1}^{n}\sum_{j=1}^{J}\rho_{\tau}\left(\bm{s}^{T}\bm{y}_{i}(t_{j})-\bm{x}_{i}^T\bm I_p \otimes \bm H^T(t_j) \bm B(\bm s)\right).
\end{equation}
In general, $\bm{\beta}(\bm{s},t)$ is a smooth function of $t$; see \cite{zhu2012multivariate} and \cite{Zhou2015qr}. To achieve the desired smoothness and also to avoid possible overfitting of the model, we impose an $L_2$ type penalty on $\bm{\beta}(\bm{s},t)$, 
\begin{equation}\label{penalty}
\textbf{P}_{\lambda}(\bm{\beta}(\bm{s},t))=\lambda\sum_{k=1}^p\int\left(\frac{\partial^2{\beta}_k(\bm{s},t)}{\partial t^2}\right)^2dt, 
\end{equation}
where $\lambda$ is the tuning parameter to control the smoothness of $\bm{\beta}(\bm{s},t)$.   The penalty in \eqref{penalty} can be written as 
\begin{equation}\label{penalty1}
\textbf{P}_{\lambda}(\bm{\beta}(\bm{s},t))=\bm B^T(\bm s)\bm \Omega\bm B(\bm s), 
\end{equation}
where $\bm \Omega = \int \left(\partial^2\bm H(t)/\partial t^2\right)^T \left(\partial^2\bm H(t)/\partial t^2\right)dt$.  Therefore, the penalized 
objective function of \eqref{bqsvcm} is of the form
\begin{equation}\label{bqsvcm1}
\bm L_{pb}(\bm y,\bm x,\bm s;\bm B(\bm s)) =\bm L_b(\bm y,\bm x,\bm s;\bm B(\bm s))+\bm B^T(\bm s)\bm \Omega\bm B(\bm s).
\end{equation}

\subsection{Multistage estimation procedure}\label{estimate}
In reality $\bm B(\bm s)$ may be continuous or piecewise continuous in terms of direction $\bm s$. 
For example, in the case $\bm Y(t) = \bm X^T\bm \beta(t) +\bm \epsilon (t)$, where $\bm \epsilon (t)$ follows independent standard normal distributions for all $t$, we have $\bm \beta_\tau(\bm s, t) = \bm s^T\bm \beta(t)$ for $\tau=.5$, which is a continuous function of the direction $\bm s$.  To estimate $\bm B(\bm s)$ using \eqref{bqsvcm1} for each individual direction $\bm s$ without considering the correlations between different directions will lose efficiency and may not be able to capture the possible continuity of $\bm B(\bm s)$.  The problems could become more severe when the sample size is limited while many directions are considered. To improve the efficiency and warrant certain smoothness, 
we can estimate $\bm B(\bm s)$ for all selected directions simultaneously.  For example, a weighted loss function may be considered, 
\begin{equation}\label{wloss}
\sum_{r=1}^d \bm L_{pb}(\bm y,\bm x,\bm s_r;\bm B(\bm s_r)) w(\bm s_r),
\end{equation}
where $\bm s_r \in \bm S = \{\bm s_r: ~r=1,\cdots,d\}$ are selected directions and $w(\bm s_r)$ are weights to characterize the corrections of the loss functions at different directions. Similar strategies have been used in \cite{bradic2011penalized}, \cite{zhao2014efficient} and others.  However, the computational complexity raised in  \eqref{wloss} by simultaneously  considering possibly hundreds of directions may be beyond the limit of the 
current computational capacity. This can be seen more clearly by noticing that the dimension of \{$\bm B(\bm s_r):~\bm s_r \in \bm S\}$ is of $dpM$, where both $d$ and $M$ are in the magnitude of hundreds.  To estimate parameters of such high dimensions, the inverse of large matrices will be involved. Unless there is certain particular structure in the matrices, in general the inverses will be computationally difficult or even impossible. 

In this  section, we propose a multistage estimation procedure based on the propagation-separation~(PS) approach \citep{polzehl2000adaptive, polzehl2006propagation} to gradually borrow information from nearby directions with increasing number of directions. 
The PS method is capable of handling the computational complexity raised by simultaneously considering multiple directions to efficiently estimate varying coefficients while guaranteeing certain smoothness along directions. To the best of our knowledges, this is the first method to construct the directional quantile regression envelopes 
by simultaneously considering multiple directions. Our multistage estimation procedure includes three main stages, namely, {\sl Stage I:  initialization, Stage II: adaptive updating, } and {\sl Stage III: stop checking}. We first briefly describe the three stages in the following. 
\begin{description}
\item[ ]  {\sl Stage I:  initialization.} Use \eqref{bqsvcm1}  to obtain the initial estimates for $\bm B(\bm s_r)$, denoted by $\hat{\bm B}_0(\bm s_r)$ for each individual direction $\bm s_r \in \bm S$.
\item[ ]  {\sl Stage II: adaptive updating.}  Use PS to adaptively update $\hat{\bm B}_{c-1}(\bm s_r)$ from the $c-1$ step for each direction in $\bm S$ by gradually increasing the number of nearby directions. 
\item[] {\sl Stage III: stop checking}.  Calculate the stopping criterion for the updated $\hat{\bm B}_{c}(\bm s_r)$ and determine if the updating needs stop.
\end{description}
The multistage estimation procedure will iterate between {\sl Stage II} and {\sl Stage III} until stop or the maximum number of iteration is reached. By gradually increasing the number of nearby directions and adaptively updating, the coefficients $\bm B(\bm s_r)$ can be efficiently  estimated with certain smoothness while the computation can be largely reduced. 

In {\sl Stage I:  initialization}, to minimize  \eqref{bqsvcm1},  it is a convex optimization problem with a nonsmooth quantile loss function and an $L_2$ type penalty in the objective function, which can be reformulated and 
effectively solved by ADMM \citep{boyd2011distributed} as shown in the next Section. The minimization of a nonsmooth quantile loss plus an $L_2$ penalty is a recurring theme in our multistage estimation procedure.  In  {\sl Stage II}, the optimization is essentially the same minimization, which will be shown next. Thanks to the computational efficiency of ADMM, our multistage estimation procedure can adaptively update the coefficients effectively. 
 
 In  {\sl Stage II: adaptive updating}, to adaptively update the estimates of  $\bm B(\bm s_r)$, we first define a nearby direction set sequence. For simplicity, we assume the directions $\bm S = \{\bm s_r: ~r=1,\cdots,d\}$ are equally distributed in $\bm S^1$ and denote the distance by $d_0$.  Given a direction 
 $\bm s_{r_0} \in \bm S$, we define a nearby direction set sequence $\{R_c(\bm s_{r_0})\}_{c=1}^C$ through a nondecreasing bandwidth sequence $\{h^c\}_{c=1}^C$, where $C$ 
 is the preselected maximum steps of iteration. In particular,  $R_c(\bm s_{r_0}) = \{\bm s_r:~\| \bm s_r-\bm s_{r_0}\|\leq d_0h^c,~\bm s_r\in \bm S\}$, where $\|\cdot\|$ denotes the $L_2$ norm. In this article, we choose $d=100$, $h=1.15$ and $C=5$. For fixed $\bm s_{r_0}$, in each iteration we update the estimates by minimizing the following loss function
 \begin{align}\label{bqsvcm2}
&\bm L_{wpb}^c(\bm y,\bm x,\bm s_{r_0};\bm B(\bm s_{r_0})) \notag\\&=\sum_{\bm s_r \in R_c(\bm s_{r_0})}w\left(\hat{\bm B}_{c-1}(\bm s_{r_0}),\bm s_r\right)\bm L_b(\bm y,\bm x,\bm s_r;\bm B(\bm s_{r_0}))+\bm B^T(\bm s_{r_0})\bm \Omega\bm B(\bm s_{r_0}), 
\end{align}
 where $\hat{\bm B}_{c-1}(\bm s_{r_0})$ are the estimates from the $c-1$ step and $w\left(\hat{\bm B}_{c-1}(\bm s_{r_0}),\bm s_r\right)$ are weights that determine the amount of information borrowed from nearby directions. Plugging the equation \eqref{bqsvcm0},  it can be seen that \eqref{bqsvcm2} is also  a nonsmooth quantile loss plus an $L_2$ penalty, which can be solved by ADMM. 
 
 The weight function $w\left(\hat{\bm B}_{c-1}(\bm s_{r_0}),\bm s_r\right)$ depends on two quantities: the distance between the directions $\bm s_{r_0}$ and $\bm s_r$ and the similarity between $\hat{\bm B}_{c-1}(\bm s_{r_0})$ and $\hat{\bm B}_{c-1}(\bm s_{r})$.  Let 
 \begin{equation}\label{dfe}
 D\left(\hat{\bm B}(\bm s_{r_0}),\hat{\bm B}(\bm s_{r})\right) = \left(\hat{\bm B}(\bm s_{r_0})-\hat{\bm B}(\bm s_{r})\right)^T\hat{\Sigma}^{-1}\left(\hat{\bm B}(\bm s_{r_0})\right)\left(\hat{\bm B}(\bm s_{r_0})-\hat{\bm B}(\bm s_{r})\right),
 \end{equation}
 where $\hat{\Sigma}\left(\hat{\bm B}(\bm s_{r_0})\right)$ is the estimated covariance matrix of $\hat{\bm B}(\bm s_{r_0})$. 
The weight function is of the form
 \begin{equation}\label{wt}
w\left(\hat{\bm B}_{c-1}(\bm s_{r_0}),\bm s_r\right) = K_{loc}\left(\|\bm s_r-\bm s_{r_0}\|/(d_0h^c)\right)K_{st}\left(D\left(\hat{\bm B}_{c-1}(\bm s_{r_0}),\hat{\bm B}_{c-1}(\bm s_{r})\right)/C_n\right),
\end{equation}
where both $K_{loc}$ and $K_{st}$ are nonnegative kernel function with compact support and $C_n$ is a tuning parameter depending on $n$. 
In this article, we choose $C_n=n^{\alpha}\chi_{1}^2(.8), \alpha\in [.3,1.3]$ as suggested  in \citet{li2011multiscale, zhu2014spatially}.  The kernel function $K_{loc}$ gives less weight to those directions far from $\bm s_{r_0}$.  The kernel $K_{st}$ downweights the directions $\bm s_r$ which has large $ D\left(\hat{\bm B}(\bm s_{r_0}),\hat{\bm B}(\bm s_{r})\right) $.  We choose $K_{loc}(u) = (1-u)_+$ and  $K_{st}(u)=\min\left(1,2(1-u)_+\right)$ in our simulations and real data analysis. For other available kernel functions, see \cite{polzehl2000adaptive, polzehl2006propagation}, \citet{li2011multiscale}, \citet{zhu2014spatially}.

 In {\sl Stage III: stop checking}, we start to check the stopping criterion after a few iterations of {\sl Stage II}, say, $c_0$ iterations. The stopping criterion is based on a normalized distance between $\hat{\bm B}_{c}(\bm s_r)$ and $\hat{\bm B}_{c_0}(\bm s_r)$, that is 
  \begin{equation}\label{dfe1}
 D\left(\hat{\bm B}_c(\bm s_r),\hat{\bm B}_{c_0}(\bm s_r)\right) = \left(\hat{\bm B}_{c}(\bm s_{r})-\hat{\bm B}_{c_0}(\bm s_r)\right)^T\hat{\Sigma}^{-1}\left(\hat{\bm B}_{c_0}(\bm s_{r})\right)\left(\hat{\bm B}_c(\bm s_{r})-\hat{\bm B}_{c_0}(\bm s_r)\right).
 \end{equation}
 The iteration stops if $\hat{\bm B}_{c}(\bm s_r)$ falls outside the ellipsoid $\{ \hat{\bm B}_{c}(\bm s_r):~D\left(\hat{\bm B}_c(\bm s_r),\hat{\bm B}_{c_0}(\bm s_r)\right) \leq C_s\}$, where $C_s$ is a preselected constant, say $C_s=\chi_{q}^{2}(.8/c)$, where $q=dpM$ is the dimension of $\hat{\bm B}_{c}(\bm s_r)$; see \citet{li2011multiscale, zhu2014spatially}. If $\hat{\bm B}_{c}(\bm s_r)$ is lying within the ellipsoid, we set $c=c+1$ and continue  {\sl Stage II} to update it in the direction $\bm s$. In general, the initial estimates are consistent, so the updated smoothing estimates shall not be too far away from the initial ones. An alternative stopping criterion is to check each individual component of $\hat{\bm B}_{c}(\bm s_r)$. Let 
   \begin{equation}\label{dfe2}
 d\left(\hat{B}^k_c(\bm s_r),\hat{B}^k_{c_0}(\bm s_r)\right) = \left(\hat{B}^k_{c}(\bm s_{r})-\hat{B}^k_{c_0}(\bm s)\right)^2\hat{\sigma}^{-2}\left(\hat{B}^k_{c_0}(\bm s_{r})\right), 
 \end{equation}
 where $\hat{B}^k_{c_0}(\bm s_{r})$ is a component of $\hat{\bm B}_{c}(\bm s_r)$ and $\hat{\sigma}^{2}\left(\hat{B}^k_{c_0}(\bm s_{r})\right)$ is the estimated variance of $\hat{B}^k_{c_0}(\bm s_{r})$. The iteration stops if any $ d\left(\hat{B}^k_c(\bm s_r),\hat{B}^k_{c_0}(\bm s_r)\right) >\chi_{1}^{2}(.8/c)$; otherwise we set $c=c+1$ and continue the {\sl Stage II}. Our simulation studies show that the latter works slightly better; so we choose to use the individual criterion in this article. 

\subsection{ADMM Algorithm}\label{admm}
The optimization problems in equations \eqref{bqsvcm1} and \eqref{bqsvcm2} are convex, which can be solved by many generic convex optimization techniques, for example, the simplex method \citep{koenker2005quantile} and the interior 
point method  \citep{koenker1996interior}. However, these methods do not take advantage of the special structure of \eqref{bqsvcm1} and \eqref{bqsvcm2}, that is, both can be split into two sub-convex optimization problems: a nonsmooth quantile loss function plus an $L_2$ type penalty.  A more efficient algorithm to solve our convex optimization problems is the alternating direction method of multipliers (ADMM). 
The ADMM algorithm was developed in the 1970s with roots in the 1950s \citep{hestenes1969multiplier, gabay1976dual},  and received renewed interest due to that
it is efficient to tackle large scale problems  and can solve optimization problems with multiple nonsmooth terms in
the objective function \citep{boyd2011distributed, Lin2013}. 
ADMM  is a powerful algorithm for  convex problems that can be decomposed into several sub-convex problems \citep{boyd2011distributed}. 
In this section, we reformulate our optimization problems of \eqref{bqsvcm1} and \eqref{bqsvcm2}  and derive their ADMM algorithms based on the observation that they can be split into two sub-convex optimization problems and one of the sub-problems has a nonsmooth function.

Let $f(x)$ and $g(z)$ be convex functions of two vectors $x$ and $z$, respectively. Suppose $A$ and $B$ are two unknown matrices and $c$ is a known vector. We have the following convex optimization problem with constraints,  
\begin{align}\label{admm1}
	{\rm  minimize}~~~~~& f(x)+g(z)   \notag\\
	s.t.~~~~~&Ax+Bz=c, 
\end{align}
which can be easily solved by ADMM. 
The augmented Lagrangian function \citep{powell1967method, hestenes1969multiplier} of \eqref{admm1} can be written as
\begin{equation} \label{admm2}
	L_{\rho}(x,z,y)=f(x)+g(z)+y^{T}(Ax+Bz-c)+(\rho/2)\|Ax+Bz-c\|^2,   
\end{equation}
where $\|\cdot\|$ is the $L_2$ norm and $\rho$ is a tuning parameter, which is chosen to be $1.2$ in this article \citep{boyd2011distributed}. Letting $u=(1/\rho)y$ and $u_k=(1/\rho)y^{k}$, the scaled augmented Lagrangian function is of the form 
\begin{equation} \label{admm3}
	L_{s\rho}(x,z,y)=f(x)+g(z)+(\rho/2)\|Ax+Bz-c+u\|^2-(\rho/2)u_2. 
\end{equation}
The ADMM algorithm finds the optimal solution of \eqref{admm3} by iterating through the following three steps:
\begin{align}
	x^{k+1}&={\rm argmin}_{x}\left(f(x)+(\rho/2)\|Ax+Bz^{k}-c+u^{k}\|^2\right),   \notag\\
	z^{k+1}&= {\rm argmin}_{z}\left(g(z)+(\rho/2)\|Ax^{k+1}+Bz-c+u^{k}\|^2\right),\notag\\
	u^{k+1}&=u^{k}+\left(Ax^{k+1}+Bz^{k+1}-c\right). 
\label{eqAdmm2}
\end{align}
Note the formulas are in the scaled form of ADMM which is often shorter and more convenient to solve than in the unscaled form, so we will use the scaled form in this article. Step one optimizes over $x$ and step two optimizes over $z$. In the last step, it brings $x$ and $z$ together to match the constraints. The key requirement for \eqref{admm1} is that $x$ and $z$ do not share common elements. In general, steps one and two in \eqref{eqAdmm2} admit simple forms, which will be illustrated later in our proposed method; for more examples, see \citet{boyd2011distributed}. 

Both \eqref{bqsvcm1} and \eqref{bqsvcm2} can be written as the sum of a nonsmooth quantile loss function and an $L_2$ type penalty. That is, 
\begin{equation}\label{admm4}
\rho_{\tau}(y - Xb) +\lambda b^T\Omega b,
\end{equation}
where $\rho_{\tau}$ is the quantile loss, $y$ is a known vector, $X$ and $\Omega$ are known matrices, $\lambda$ is the tuning parameter and $b$ is 
the vector we optimize over. To adapt to the ADMM algorithm, we reformulate \eqref{admm4} to 
\begin{align}\label{admm_qr}
	{\rm  minimize}~~~~~&\rho_\tau(r)+\lambda b^T\Omega b,  \notag\\
	s.t.~~~~~&r+Xb = y. 
\end{align}
Observing that $f(r) = \rho_\tau(r)$, $g(b) = \lambda b^T\Omega b$, $A = I$, $B = X$, and $c=y$ comparing with \eqref{admm1}, similar to 
\eqref{eqAdmm2} we solve \eqref{admm_qr} by iterating the following 
\begin{align}
	r^{k+1}&={\rm argmin}_{r}\left(\rho_\tau(r)+(\rho/2)\|r+Xb^{k}-y+u^{k}\|^2\right),   \notag\\
	b^{k+1}&= {\rm argmin}_{b}\left(\lambda b^T\Omega b+(\rho/2)\|r^{k+1}+Xb-y+u^{k}\|^2\right),\notag\\
	u^{k+1}&=u^{k}+\left(r^{k+1}+Xb^{k+1}-y\right).
\label{eqAdmm3}
\end{align}
The first step in  \eqref{eqAdmm3} can be simplified by the soft-thresholding operator. That is, step one has the following closed form,  
\begin{equation}\label{admm_s1}
r^{k+1}=S_{1/(2\rho)}\left(u^k-y+Xb^k-(2\tau-1)/(2\rho)\right),
\end{equation}
where  $S_{a}(v)=(v-a)_{+}-(-v-a)_{+}$ is a soft-thresholding operator. The second step is a least square loss plus an $L_2$ type penalty and admits 
the following ridge regression type closed form, 
\begin{equation}\label{admm_s2}
b^{k+1}=\left(2\lambda\Omega/\rho+X^TX\right)^{-1}X^T\left(y-r^{k+1}+u^k\right).
\end{equation}
Note that the inverse of $2\lambda\Omega/\rho+X^TX$ only needs to be calculated once. Therefore, the iterations in \eqref{admm_qr} are vary fast and efficient. 

We use the termination criterion suggested by \citet{boyd2011distributed}, which is based on primal residuals $r_{pri}$ and dual residuals $r_{dual}$. At the $k$-th iteration, the primal residuals $r^{k}_{pri}$ and dual residuals $r^{k}_{dual}$
 are calculated according to 
 \begin{align}
	r_{pri}^{k}&=y-Xb^{k}-r^{k},   \notag\\
r_{dual}^{k}&=\rho X(b^{k}-b^{k-1}),
\label{eqAdmmr}
\end{align}
respectively. The termination criterion is 
\begin{equation}\label{admm_t}
\|r_{pri}^{k}\|\leq \epsilon^{pri}~ \mbox{and}~ \|r_{dual}^{k}\| \leq\epsilon^{dual},
\end{equation}
where $\epsilon^{pri}>0$ and $\epsilon^{dual}>0$ are feasibility tolerances for the primal and dual feasibility conditions. These tolerances can be chosen using an absolute and relative tolerances \citep{boyd2011distributed},
\begin{align}\label{admm_tol}
	\epsilon^{pri}&=\sqrt{p}\epsilon^{abs}+ \epsilon^{rel}\max\left(\|r^{k}\|_{2},\|Xb^k\|,\|y\|\right), \notag\\ 
	\epsilon^{dual}&=\sqrt{n}\epsilon^{abs}+\epsilon^{rel}\|u^k\|, 
\end{align}
where $\epsilon^{abs}>0$ and $\epsilon^{rel}>0$ are absolute and relative tolerance, respectively. 
The absolute tolerance and relative tolerance can be any small numbers in practical calculations. For example, we choose  $\epsilon^{abs}=10^{-4}$ and $\epsilon^{rel}=10^{-2}$ in this article.

\section{Numerical Studies}

\subsection{Simulation Studies}\label{sim}
In this section, we conduct simulation studies to evaluate the performance of our proposed methods. 
We investigate models of different types of coefficients and  various error distributions. 
Let $\bm{Y}(t)=(Y_{1}(t),Y_{2}(t))^T\in\real^2$ be a bivariate functional response at time $t$, where $t\in [0,1]$, $\bm{X}\in \real^3$ be the covariates of interest such that $\bm{X}=(1,X_{1},X_{2})^T$; and they have the following relationship,
\begin{equation}\label{model}
	\bm Y(t)=\left(\bm{X}^T\bm{\beta}_1(t), \bm{X}^T\bm{\beta}_2(t)\right)^T+\bm \epsilon(t), 
\end{equation}
where 
$\bm{\beta}_i(t)=\left(\beta_{i0}(t),\beta_{i1}(t),\beta_{2i}(t)\right)^{T}$, for $i=1$ and $2$.  
In our simulation studies,  we set $ X_{1}\sim Bernoulli~(.5), $ and $X_{2}\sim Uniform(0,1)$, where the  choice  of two variables is motivated by our DTI data in Section \ref{real}. In general, $X_1$ and $X_2$ represent binary (e.g. gender or diagonal status) and scaled continuous variables (e.g. age, or height). Two types of varying coefficients $\bm{\beta}(t)$ are considered, smooth and rough, which are 
\begin{align}\label{coeff1}
\bm \beta_1(t) &= \left( 2t+1, \sin (t)+2, \cos (t)-2\right)^T, \notag\\ 
\bm \beta_2(t) &= \left( 2t-1, \cos(t)-2, \sin (t)+3 \right)^T,
\end{align}
and 
\begin{align}\label{coeff2}
\bm \beta_1(t) &= \left( 40t/(2t+1), (t^2+3)/(t-2), t+3\right)^T, \notag\\ 
\bm \beta_2(t) &= \left( \log(t+1), t+1, 3t^2-2 \right)^T, 
\end{align}
respectively. 
Furthermore, we  consider  three types of error distributions, namely, normal, $t$,
 and $\chi^2$ distributions. In particular,  the three error distributions are 
\begin{description}
\item[ ] (I) a bivariate normal distribution 
\begin{equation}\label{e1}
\bm \epsilon(t) \sim N(\bm \mu,\Sigma),
\end{equation}
where $\bm \mu = (0,0)^T$ and $\Sigma =\rm diag(.8,.8)$;
\item[ ] (II) a bivariate $t$ distribution with degrees of freedom $3$
\begin{equation}\label{e2}
\bm \epsilon(t) \sim t_3(\bm \mu,\Sigma),
\end{equation}
where $\bm \mu = (0,0)^T$ and $\Sigma =\rm diag(.8^5,.8^5)$; and 
\item[ ] (III) a bivariate $\chi^2$ distribution with degrees of freedom $3$
\begin{equation}\label{e3}
\bm \epsilon(t) \sim .8\left(a_{1}^2+a_{2}^2+a_{3}^2,~a_{3}^2+a_{4}^2+a_{5}^2\right)^T,
\end{equation}
where $a_i\sim N(0,1)$ for $i=1,\cdots,5$ are mutually independent. 
\end{description}
The first error distribution is very common; the second distribution mimics heavy tailed distributions and the last one stands for skewed and correlated distributions. We carefully choose the parameters of the error distributions so that the models have comparable signal-to-noise ratios (SNRs). 

We choose $J = 50$ equally spaced points t from the interval [0,1] and the sample size of our Monte Carlo simulations is set to be  $200$.
Then we estimate the coefficients using our proposed methods in Section \ref{estimate}.  In particular, we choose $100$ evenly spaced directions  in [$-\pi, \pi$] and B-spline basis with $14$ evenly spaced knots in $[.02,  .93]$.  The ADMM algorithms in Section \ref{admm} are implemented in R with core parts written in C. Then we construct the $\tau = \{.05, .1,.2\}$-th directional  quantile envelopes according to \eqref{qe_dfn1}; for detailed construction algorithms, see \citet{kong2009multivariate, hallin2010multivariate}, and \citet{kong2012quantile}.
To evaluate the resulting directional envelopes, we look at two measures, namely, the  envelope curvature --- the average change rate of slope of the envelope, and the coverage rate --- the proportion of data points covered by the envelope, denoted by $\kappa$ and $\nu$, respectively. 
We repeat our Monto Carlo simulations $100$ times and calculate the  mean and standard deviation of $\kappa$ and $\nu$ of the $100$ replications from the initial and updated estimates.  The initial varying coefficients are calculated by \eqref{bqsvcm1} and the updated ones are estimated by our multistage estimation procedure. 

%%%%%%%%%%%%%%%%%%%%%%%% Table 1 %%%%%%%%%%%%%%%%%%%%%%%
\setlength{\tabcolsep}{.1cm}
\begin{table}[ht!]
\caption{The envelope curvatures $\kappa$ and coverage rates $\nu$ of the true, initial, and updated directional quantile envelopes at quantile levels $\tau = \{.05,.1,.2\}$ for simulated models with smooth (S) or rough (R) coefficients  and different error distributions, I, II, and III. The standard deviations of the initial and updated $\kappa$ and $\nu$ are listed in the brackets. }
\vspace*{12pt}
\centering
	\begin{tabular}{|c|c|c||ccc|ccc|ccc|}
		\hline
		\multicolumn{3}{|c||}{}
		& \multicolumn{3}{|c|}{$\tau$=.05} & \multicolumn{3}{|c|}{$\tau$=.1}
		& \multicolumn{3}{|c|}{$\tau$=.2} \\\cline{4-12}
		\multicolumn{3}{|c||}{} &True & Initial & Updated &True & Initial & Updated & True&Initial & Updated \\\hline\hline
		\multirow{6}*{S}&\multirow{2}*{I}
		& $\kappa$ &.80&  3.0(2.2) & 1.5(.9) &1.08& 4.28(3.9) & 1.78(.9) &2.63& 6.03(3.8) & 4.13(2.1) \\
		&& $\nu$&.740 & .742(.024) & .741(.013) &.560& .589(.020) & .575(.020) &.295&  .298(.017) &  .295(.016) \\\cline{2-12}
		 &\multirow{2}*{II}
		& $\kappa$ &1.37&  2.97(2.0) & 2.27(2.0) &2.02&  4.0(2.0)  & 2.82(2.2) &2.34& 9.94(10.2) & 6.54(4.4)\\
		&&$\nu$ &.790& .768(.026) & .776(.013) &.620& .623(.023) & .621(.023)&.340& .328(.020)&   .340(.018) \\\cline{2-12}
		&\multirow{2}*{III}
		& $\kappa$ &.56& .65(1.2) & .64(.9)&1.23&  2.43(1.9) & 2.03(1.8)&1.86& 3.56(2.0) & 2.76(1.3)\\
		&& $\nu$ &.723& .614(.017) & .722(.019) &.540& .512(.022)& .551(.018) &.280& .312(.015)&  .287(.014) \\\hline\hline
				\multirow{6}*{R}&\multirow{2}*{I}
		& $\kappa$ &.97&  2.77(2.5) & 2.17(1.7) &2.66& 4.68(2.4) & 2.72(2.1) &3.29& 6.19(5.5) & 3.79(2.9) \\
		&& $\nu$&.740 & .802(.022) & .740(.016) &.560& .600(.025) & .533(.021) &.295&  .301(.018) &  .293(.017) \\\cline{2-12}
		 &\multirow{2}*{II}
		& $\kappa$ &1.52&  2.82(1.8) & 1.53(.9) &1.50&  4.10(2.7)  & 4.00(1.8) &2.96& 6.86(3.6) & 5.06(3.3)\\
		&&$\nu$ &.790& .809(.019) & .792(.010) &.620& .628(.024) & .626(.018)&.340& .312(.018)&   .333(.017) \\\cline{2-12}
		&\multirow{2}*{III}
		& $\kappa$ &.90& 1.10(1.5) & 1.60(1.5)&1.01&  1.71(1.2) & 1.71(1.1)&3.17& 3.97(3.1) & 3.18(2.2)\\
		&& $\nu$ &.720& .682(.022) & .719(.017) &.540& .583(.019)& .570(.018) &.280& .317(.014)&  .284(.013) \\\hline
	\end{tabular}
\label{sim_tab}
\end{table}

%%%%%%%%%%%%%%%%%%%% FIGURE 1 %%%%%%%%%%%%

\begin{figure}[ht!]
\includegraphics[width=.95\textwidth] {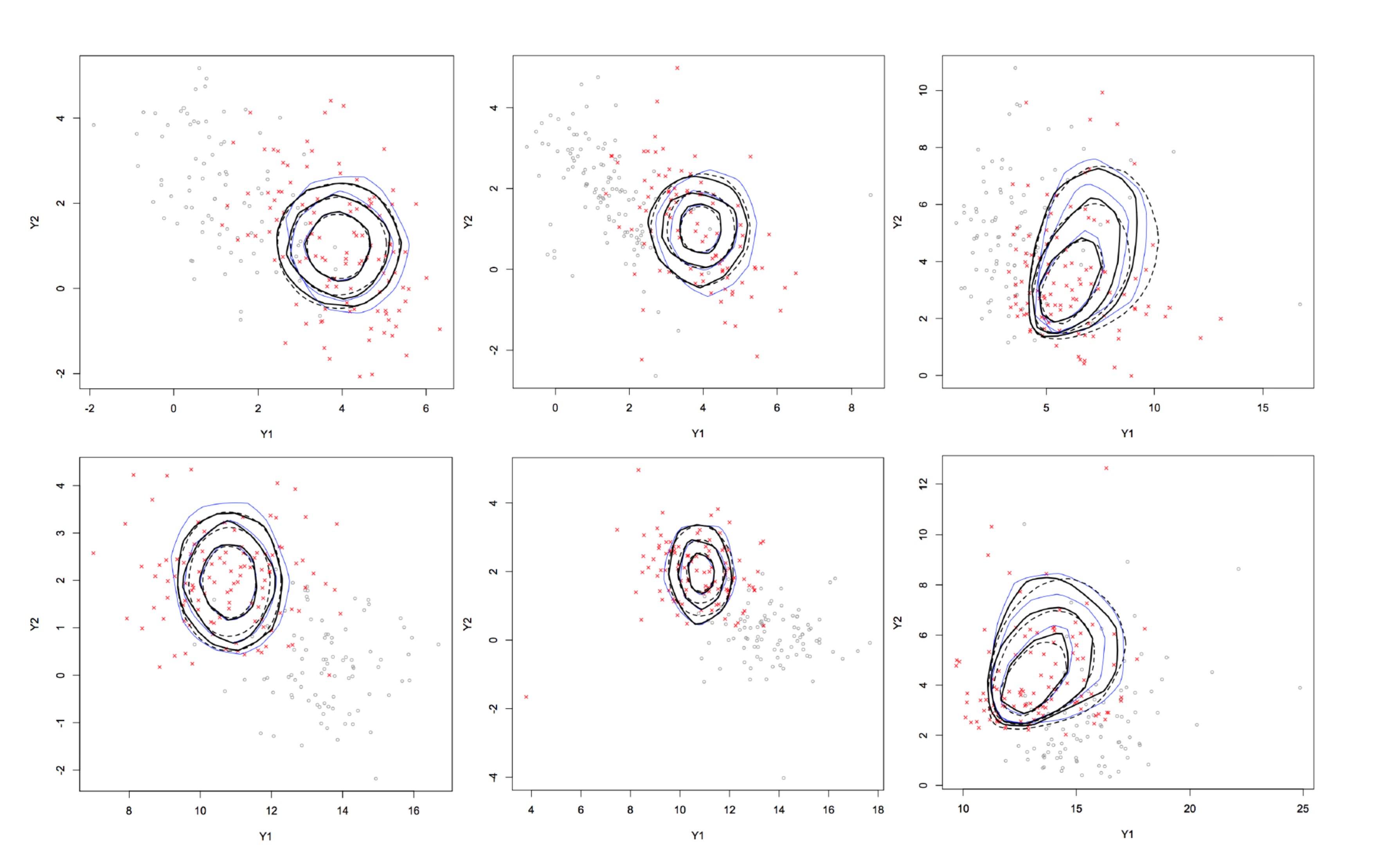}  
\caption{One selected data set at $X_1 = 0$ (grey circle) or $X_1 = 1$ (red cross),  $X_2=.5$, and around $t=.7$ overlaid the true (thick dashed dark), initial (thin solid blue), and the updated (thick solid dark) directional envelopes at the data point $(X_1,X_2,t) = (1,.5,.7)$ from the models with smooth (upper panels) or rough (lower panels) coefficients and three error distributions I (left panels), II (middle panels), and III (right panels).}
\label{sim_fig}
\end{figure}

We present the results in Table \ref{sim_tab} for a selected data point with $X_{1}=1,X_{2}=.5$ at $t=.7$, where the true $\kappa$ and $\nu$ values are calculated from the direction quantile envelopes from $5000$ generated observations at the selected point in our simulation models. For the models with smooth coefficients, Table \ref{sim_tab} shows that  the envelope curvature $\kappa$ values for the updated quantile envelopes are closer to the true values compared with the initial ones at all the three selected quantile levels. Moreover, the updated $\kappa$ values have smaller standard deviations than the initial ones, which are shown in the bracket in Table \ref{sim_tab}. Similar patterns are observed for the models with rough coefficients in Table \ref{sim_tab} , some patterns are weaker or slightly reversing though. These indicate that our proposed multistage estimation procedure is capable of obtaining quantile envelopes with desired smoothness and smaller variations. In Table \ref{sim_tab} , for both smooth and rough coefficients we observe that the updated coverage rate $\nu$  values
have much smaller bias compared with the initial estimates and in general have smaller or comparable standard deviations, shown in the bracket in Table \ref{sim_tab}.  Therefore, our multistage estimation procedure is able to substantially reduce the bias of the coverage rate  $\nu$ by adaptively updating our estimates through borrowing information from nearby directions.
%, where the bias may be caused by finite sample sizes and the B-spline approximations.  
This implies  that the adopted PS method is effective in parameter estimation (and thus predicting data points),  as it improves not only the smoothness but also the accuracy of the regression coefficients.   In Figure \ref{sim_fig}, the true (thick dashed dark), initial (thin solid blue), and the updated (thick solid dark) directional envelopes at the selected data point  further confirm our observations that the updated envelopes are closer to the true ones and smoother than the initial ones. In summary, our proposed multistage estimation procedure is more efficient in constructing directional quantile envelopes. Moreover, it provides quantile envelopes with not only desired smoothness but also more accurate coverage rates.

We conclude this section with some comments on the choice of parameters in the multistage estimation procedure, which is crucial when applying this method, especially in the PS method. One key parameter is the scale parameter $C_n$ in the kernel function $K_{st}$ to penalize the dissimilarity between two directions in a manner similar to bandwidth in local polynomial smoothing \citep{fan1996local}.  Based on our experience, we recommend $C=n^{\alpha}\chi_{1}^2(0.8)$, where $\alpha\in [0.3,1.3]$,  $ n$ is the sample size and $\chi_{1}^2(u)$ is the $u$-th upper quantile 
of the chi-square distribution with  $1$ degree of freedom \citep{li2011multiscale, zhu2014spatially}. 
 When $\alpha$ increases, $K_{st}$ increases and more information of the nearby directions is included.
 Another important parameter  is the penalty parameter  $\lambda$ that controls the smoothness along $t$ --- if $\lambda$ is small, less smoothness is imposed; otherwise, more smoothness is imposed. We suggest choose $\lambda$ from $\{0.001, 0.01, 0.1, 1\}$. Although cross-validation can always be used to choose $\lambda$, we only do that in the initial stage for each quantile level and in the following update stages we choose the same $\lambda$ to save computation time. 

\subsection{Neuroimaging data analysis}\label{real}
The data set consists of 128 healthy full-term infants (75 males and 53 females)  from a clinical study on early brain development, which was approved by the Institutional Review Board of the University of North Carolina at Chapel Hill.  The mean gestational age at MR scanning of the 128 infants was 298 $\pm$ 17.6 days. For each subject, the DTI images were obtained by using a
single shot EPI DTI sequence (TR/TE=5400/73 msec) with eddy current
compensation.  The six non-collinear directions were applied at the
$b$-value of 1000 s/mm$^2$ with a reference scan ($b=0$). The voxel
resolution was isotropic 2 mm, and the in-plane field of view was
set at 256 mm in both directions.  To improve the signal-to-noise
ratio of the images, a total of five scans were acquired and
averaged.

\begin{figure}[ht!]
\includegraphics[width=.95\textwidth] {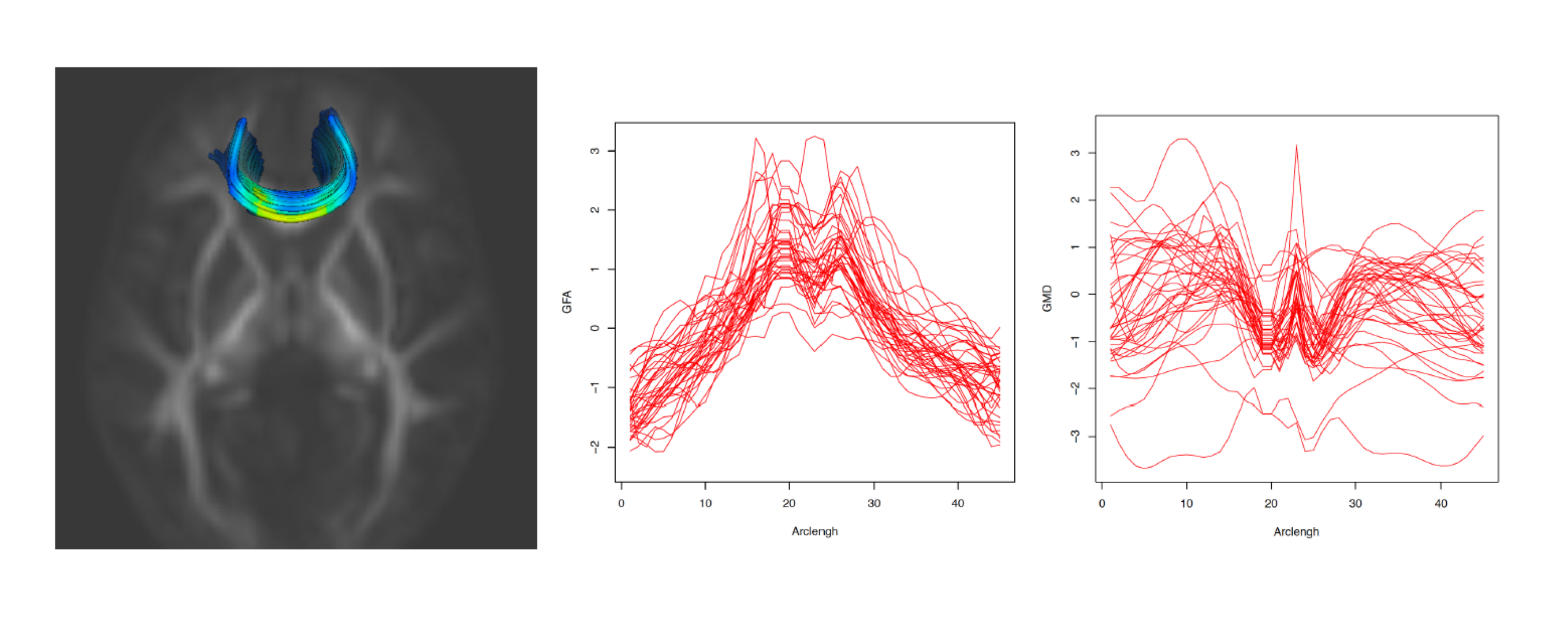}  
\caption{Genu tract (left panel) and two diffusion properties,  fractional
anistropy (GFA) (middle panel) and mean diffusivity (GMD) (right panel), observed in $45$ grid points along the genu tract from $40$ randomly selected infants.}
\label{dti_fig1}
\end{figure}

To construct the diffusion tensors, there are two key steps  including a  weighted least
squares estimation method \citep{zhu2007statistical} and a DTI atlas building process followed by an atlas-based
tractography procedure; for more details see \citet{zhu2011fadtts}.   In this article, we focus on the fiber bundle of
the genu of the corpus callosum (GCC), shown in the left panel of Figure \ref{dti_fig1}, which is an area of white matter in the
brain. Two diffusion properties, standardized fractional
anistropy (FA) and standardized mean diffusivity (MD),  are to be studied; They are bivariates functional responses of arclength observed in $45$ grid points, shown in the middle (FA along GCC - GFA) and right panels (MD along GCC - GMD) of Figure \ref{dti_fig1}.  FA and MD, respectively, measure  the
inhomogeneous extent of local barriers to water diffusion and the
averaged magnitude of local water diffusion.   

%%%%%%%%%%%%%%%%%%%% FIGURE 2 %%%%%%%%%%%%

In this analysis, our aim is to study the quantile association between two diffusion properties (FA and MD) and a set of covariates. In particular, we fit model \eqref{bqvcm1}, where $\bm Y = (\mbox{GFA},~\mbox{GMD})^T$ --- the standardized GFA and GMD values and $\bm x_i= (1, \mbox{Gender}),\mbox{G}$ with G standing for the gestational age. After finding the directional quantile coefficients at quantile levels $\tau = \{.05,.1,.2,.3\}$ using $100$ selected directions, we construct the corresponding directional quantile envelopes by \eqref{qe_dfn1}. We chose $100$ directions because previous studies have shown that $100$ directions are sufficient to characterize the envelopes \citep{kong2009multivariate,kong2012quantile}. Using the directional quantile envelopes, our analysis provides new insights on the early brain development at both population and individual levels.

%%%%%%%%%%%%%%%%%%%% FIGURE 3 %%%%%%%%%%%%

\begin{figure}[ht!]
\includegraphics[width=.95\textwidth] {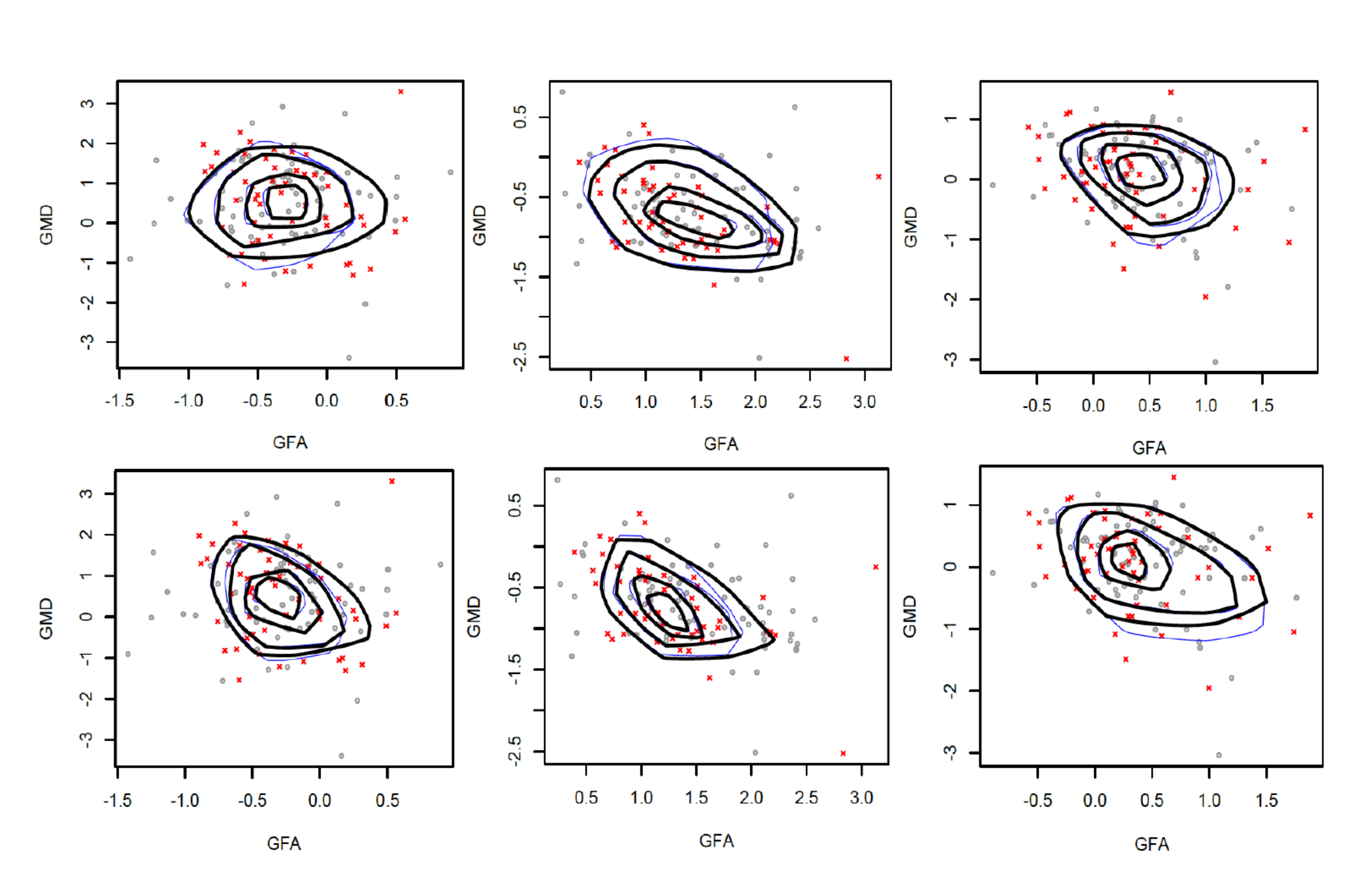}  
\caption{Quantile envelopes of GFA and GMD at gestational age~$300$ for males (upper panels) and females (lower panels) at three different arclengths, $10$ (left panels), $20$ (middle panels) and $30$ (right panels) overlaid on the observed data points (males - grey circle, females - red cross). The quantile levels from outer  to inner envelopes are $\tau = \{.05,.1,.2,.3\}$ and the thin blue and  thick dark envelopes are initial and updated envelopes, respectively.}
\label{dti_fig2}
\end{figure}

To illustration the information on population level we can obtain from the resulting envelopes, we look at the quantile envelopes of GFA and GMD at gestational age~$300$ for males (upper panels) and females (lower panels) at three different arclengths, $10$ (left panels), $20$ (middle panels) and $30$ (right panels), separately; see Figure \ref{dti_fig2}. We observe similar patterns for males and females in terms of shape and location of the quantile envelopes at different arclenghs.  On the other hand, we also observe some differences; for example, the envelope sizes for males are bigger at archlenths $10$ and $30$ while smaller at arclength $30$ than those for females. The envelope shapes and locations change with arclengh for both males and females; for instance,  the ranges of the $95$ percent quantile envelope of males are $GMD\in (-1, 2)$ and $GFA\in (-1, 0.5)$ at arclength $10$, $GMD\in (-1.5, 0.25)$ and $GFA\in (0.5, 2.5)$ at arclength $20$, and  $GMD\in (-1, 1)$ and $GFA\in (-0.25, 1.25)$ at arclenth $30$. The $90$ and $95$ initial quantile envelopes for males are crossing with each other while the updated ones are not by borrow information from nearby directions (upper middle panel in Figure \ref{dti_fig2}). This indicates that our method can effectively use the information in the data to yield more harmonious model structures.

\begin{figure}[ht!]
\includegraphics[width=.95\textwidth] {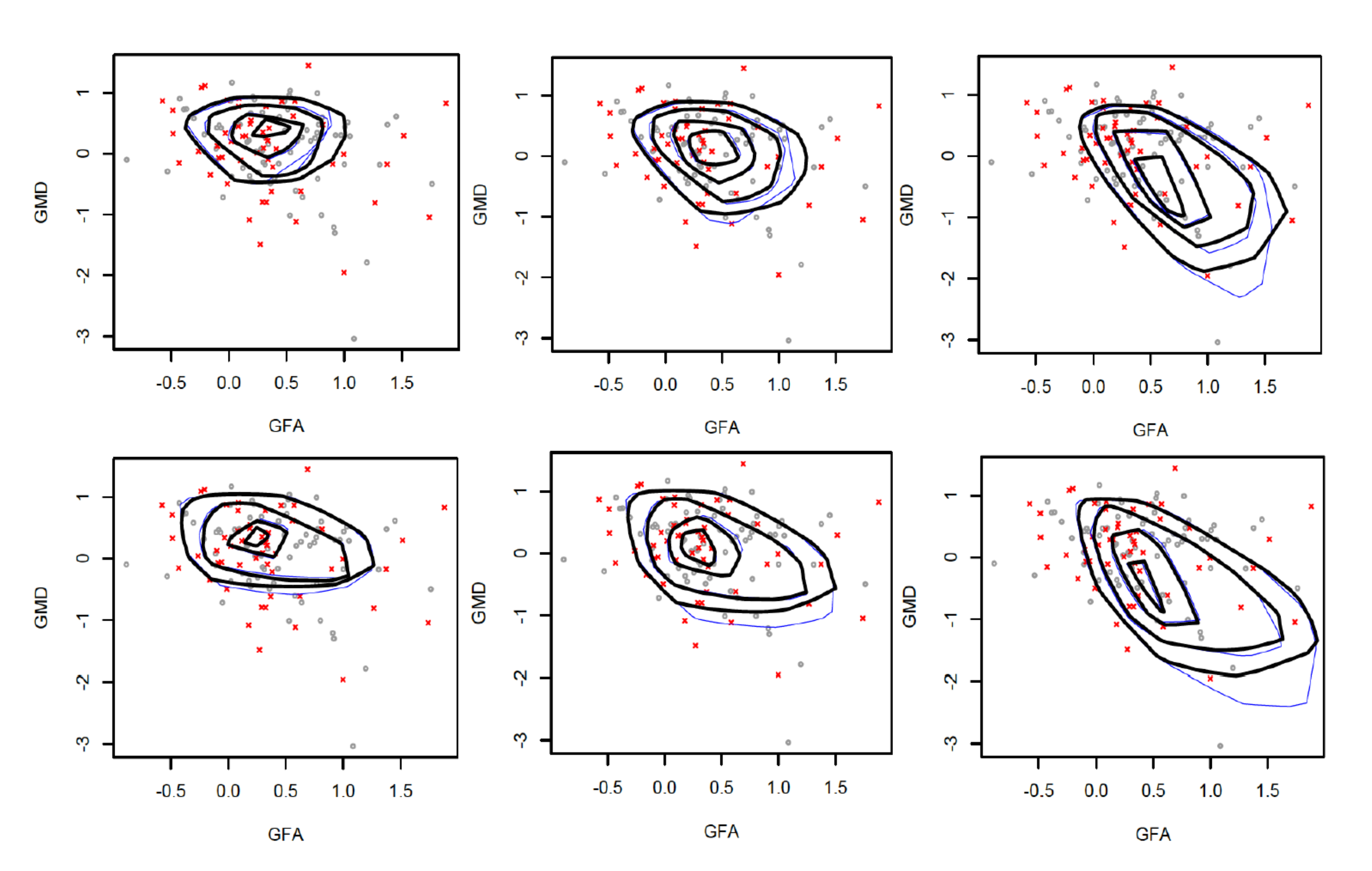}  
\caption{Quantile envelopes of GFA and GMD at arclength~$30$ for male (upper panels) and female (lower panels) at three different gestational ages, $280$ (left panels), $300$ (middle panels) and $340$ (right panels) overlaid on the observed data points (male - grey circle,  female - red cross). The quantile levels from outer  to inner envelopes are $\tau = \{.05,.1,.2,.3\}$ and the thin blue and  thick dark envelopes are initial and updated envelopes, respectively.}
\label{dti_fig3}
\end{figure}

In Figure \ref{dti_fig3}, we display quantile envelopes of GFA and GMD at arclength~$30$ for males (upper panels) and females (lower panels) at three different gestational ages, $280$ (left panels), $300$ (middle panels) and $340$ (right panels).  The quantile envelopes show consistent patterns for males and females in terms of shape and location. The quantile envelope sizes increase with gestational ages for both males and females; the sizes for males are smaller than  those for females though. As gestational age increases, the joint distributions of GFA and FMD become more skewed to the lower right corner (right panels in Figure \ref{dti_fig3}) as evidenced by the shapes of the quantile envelopes and the distances between them. 

%%%%%%%%%%%%%%%%%%%% FIGURE 4 %%%%%%%%%%%%

%%%%%%%%%%%%%%%%%%%% FIGURE 5 %%%%%%%%%%%%

\begin{figure}[ht!]
\centering
\includegraphics[width=.8\textwidth] {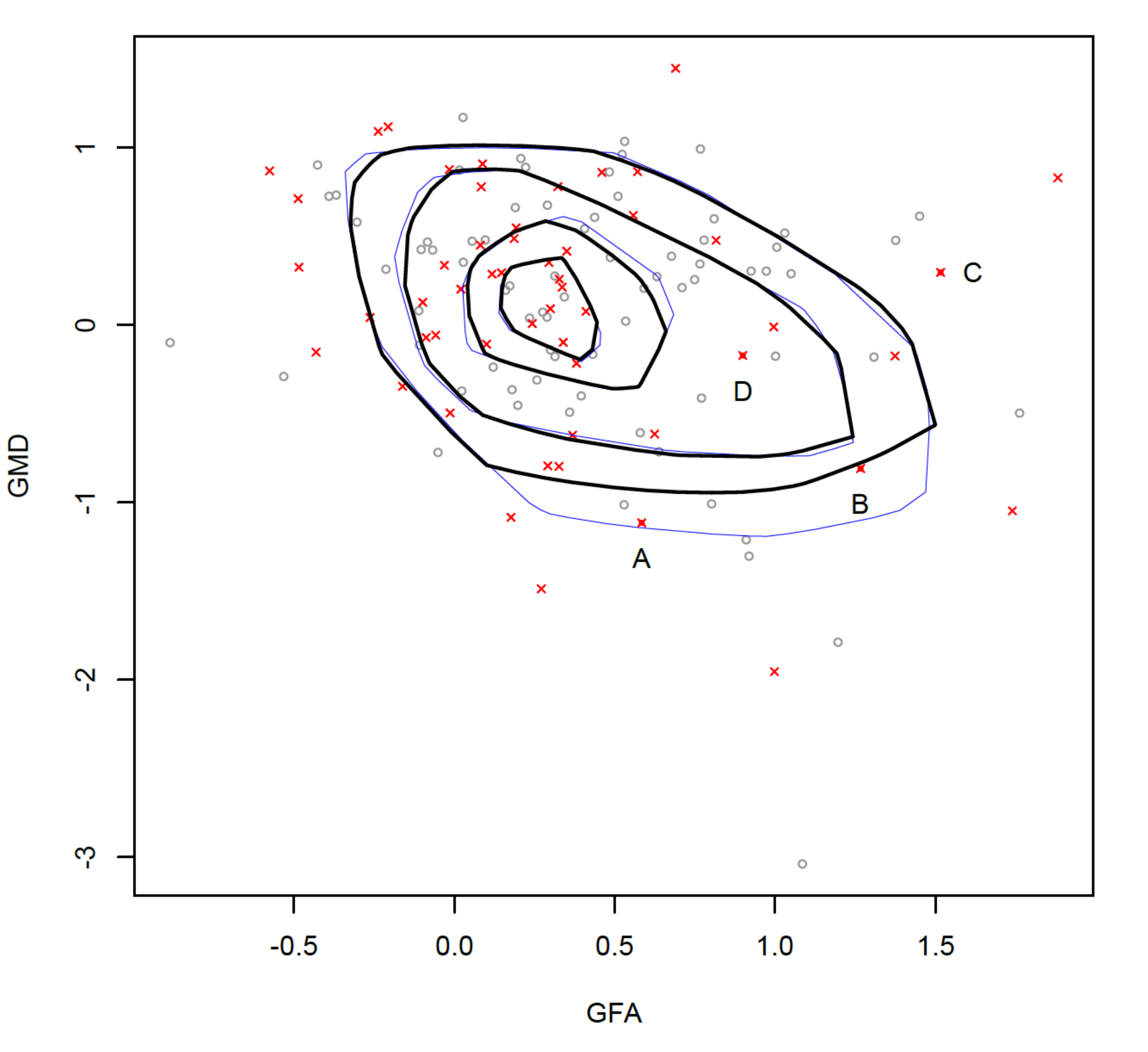}  
\caption{Quantile envelopes of GFA and GMD at arclength~$30$ for females at gestational age $300$ overlaid on the observed data points (male - grey circle, female - red cross). The quantile levels from outer  to inner envelopes are $\tau = \{.05,.1,.2,.3\}$ and the thin blue and  thick dark envelopes are initial and updated envelopes, respectively.}
\label{dti_fig4}
\end{figure}

In the end, we demonstrate how to use the directional quantile envelopes at individual levels to gain insights to early brain development. For this purpose, we display in Figure \ref{dti_fig4} the quantile envelopes of GFA and GMD at arclength~$30$ for females at gestational age $300$ overlaid on the observed data points (male - grey circle, female - red cross). The quantile levels from outer  to inner envelopes are $\tau = \{.05,.1,.2,.3\}$.  In particular, we look at four female infants, denoted by $A$, $B$, $C$, and $D$ in Figure \ref{dti_fig4}.  The infant $D$ is within the $90$ percent quantile envelope and her brain is normally developing while the infant $C$ may need further clinical investigation as she is outside the $95$ percent quantile envelope. Interestingly, we find that both infants $A$ and $B$ are lying inside the initial $95$ percent quantile envelope but are outside the updated $95$ percent quantile envelope. Therefore, it may be worth  to further conducting more clinical examinations to evaluate the brain development of these two infants as well.

\section{Discussion}\label{conclusion}
This article studies a novel estimation method in bivariate quantile varying coefficient model for functional responses based on the directional quantile concept. We approximate the  varying coefficients by the B-spline basis and impose an $L_{2}$-type penalty to achieve desired smoothness.  A multistage estimation procedure is proposed based on the PS approach to borrow information from nearby directions. The PS method is capable of handling the computational complexity raised by simultaneously considering multiple directions to efficiently estimate varying coefficients while guaranteeing certain smoothness along directions.  The proposed objective function is reformulated into a new form and then the ADMM is utilized to solve the optimization problem.  Simulation studies demonstrate that  the our proposed method is more efficient in constructing directional quantile envelopes. Moreover, it provides quantile envelopes with not only desired smoothness but also more accurate coverage rates. 
 We analyze a real data on DTI properties and our analysis yields new insights are provided on the early brain development at both population and individual levels.

There are several topics that merit further research. The asymptotic properties, such as consistency and asymptotic normality,  of  our proposed method could be developed. In particular,  the asymptotic properties of \eqref{bqsvcm1} may be studied by using the techniques in \citet{li2007quantile}. In the multistage estimation procedure, similar regular conditions in \citet{zhu2014spatially} could be imposed to pursue its asymptotic properties. To estimate the varying coefficients, other basis functions, for example, wavelet basis, could be used \citep{tsatsanis1993time,silverman2005functional}.  Another alternative is to use local kernel polynomial smoothing method \citep{fan1996local, zhu2012multivariate}.To achieve certain properties of the varying coefficients in \eqref{bqsvcm1}, other penalty functions  can be adapted; for instance, LASSO, SCAD or MCP can be used to yield sparse estimates of the B-spline basis \citep{tibshirani1996regression,fan2001variable,zhang2010nearly}.  Furthermore, our estimation method can be easily extended to multivariate quantile varying coefficient model for functional responses \citep{kong2012quantile,zhu2012multivariate} by modifying the multistage estimation procedure.

\newpage

\bibliographystyle{natbib}

\bibliography{2015_10_shu}
\end{document}